\documentclass[11pt,twoside,a4paper]{article}
\usepackage[T1]{fontenc}
\usepackage{graphicx}
\usepackage{amsmath}
\usepackage{amssymb}
\usepackage{hyperref}
\usepackage{url}
\usepackage{siunitx}
\usepackage{rotating}
\usepackage{longtable}
\usepackage{tabularx}
\usepackage{geometry}
\usepackage{pdflscape}

\title{Optical properties of two complementary samples of intermediate Seyfert galaxies}
\date{}
\author{Benedetta Dalla Barba $^{1,2,*}$, Marco Berton $^{3}$, Luigi Foschini $^{2}$, \\
Giovanni La Mura $^{4,5}$, Amelia Vietri $^{6}$, and Stefano Ciroi $^{6}$}

\begin{document}
\maketitle
{\scriptsize{\noindent $^{1}$ \quad \noindent Università degli studi dell'Insubria, 22100 Como, Italy\\
$^2$ \quad Osservatorio Astronomico di Brera, Istituto Nazionale di Astrofisica (INAF), 23807 Merate, Italy\\
$^3$ \quad European Southern Observatory (ESO), 19001 Santiago de Chile, Chile\\
$^4$ \quad Osservatorio Astronomico di Cagliari, Istituto Nazionale di Astrofisica (INAF), 09047 Selargius, Italy\\
$^5$ \quad Laboratório de Instrumentação e Física Experimental de Partículas (LIP), 1649-003 Lisboa, Portugal\\
$^6$ \quad Dipartimento di Fisica e Astronomia, Università di Padova, 35122 Padova, Italy\\
$^*$ \quad Correspondence: benedetta.dallabarba@inaf.it}}

\begin{abstract}
We present preliminary results of the analysis of optical spectra of two complementary samples of Seyfert galaxies. The first sample was extracted from a selection of the 4th Fermi Gamma-ray Large Area Telescope (4FGL) catalog, and consists of 9 $\gamma$-ray emitting jetted Seyfert galaxies. The second one was extracted from the Swift-BAT AGN Spectroscopic Survey (BASS), and is composed of 38 hard-X ray selected Active Galactic Nuclei (AGN). These two samples are complementary, with the former expected to have smaller viewing angles, while the latter may include objects with larger viewing angles. We measured emission line ratios to investigate whether the behavior of these Seyferts can be explained in terms of obscuration, as suggested by the well-known Unified Model (UM) of AGN, or if there are intrinsic differences due to the presence of jets, outflows, or the evolution. We found no indications of intrinsic differences. The UM remains the most plausible interpretation for these classes of objects even if some results can be challenging for this model.
\end{abstract}

Keywords: Seyfert galaxies; intermediate Seyfert; AGN; optical spectroscopy

\section{Introduction and motivation}
Seyfert galaxies were first introduced by Seyfert in 1943 \cite{Seyfert}, where he noted the presence of unusual broad emission lines in the optical spectra of a sample of nearby galaxies. Following this seminal work, other authors \cite{Weedman70}, \cite{Weedman73}, \cite{Khachikian71}, \cite{Khachikian74} emphasized the need to categorize Seyferts into at least two groups: Seyfert 1 and Seyfert 2 (Sy1s and Sy2s, respectively). Sy1 correspond to Active Galactic Nuclei (AGN) with low inclination or nearly face-on orientations, where the Broad-Line Region (BLR) is visible in optical spectra and produces a broad component in the permitted lines, such as the Balmer lines. In contrast, Sy2 have high inclination or nearly edge-on orientations. This orientation difference is not linked to the host galaxy properties, but to the presence of a dusty torus, which completely obscures the BLR in Sy2, allowing only the narrow lines from the Narrow-Line Region (NLR) to be visible in the spectra. This interpretation of Seyferts is based on the well-known Unified Model (UM) introduced by Antonucci \cite{Keel}, \cite{Antonucci}, \cite{Urry}. According to this model, the distinction between Sy1 and Sy2 galaxies is sharp, although there are numerous sources with intermediate properties observed to date. These sources have been classified as Intermediate Seyferts (IS), defined as Seyferts with spectral properties falling between types 1 and 2 \cite{Osterbrock76}, \cite{Osterbrock77}. In the spectra of IS galaxies, the Balmer lines exhibit a composite profile composed of a narrow peak from the NLR superimposed on a broad component from the BLR. Consequently, IS galaxies can be further subdivided based on the prominence of the BLR relative to the NLR in the permitted lines. This division results in categories such as Sy1.2, Sy1.5, Sy1.8, and Sy1.9, with the sequence representing a decreasing contribution of the broad component compared to the narrow one. The determination of an IS galaxy's type can be achieved using two methods, the first of which is: 

\begin{equation}
	R=\frac{\mathrm{[O III]}\lambda5007}{\mathrm{H}\beta_{\mathrm{broad}}}
\end{equation}

\noindent where [O III]$\lambda$5007 is the line flux of the oxygen line, the same for H$\beta$ \cite{Whittle}, \cite{Netzer}. Here we identify four classes using Table~\ref{tab1}:

\begin{table}[h!] 
\caption{Seyfert classification according to Whittle \cite{Whittle}.\label{tab1}}
\newcolumntype{C}{>{\centering\arraybackslash}X}
\begin{tabularx}{\textwidth}{C|C}
\hline
\textbf{Seyfert type}	& \textbf{Condition} \\
\hline
Sy1		&	 R $\leq$ 0.3		\\
Sy1.2	&	 0.3 < R $\leq$ 1		\\
Sy1.5 	&	 1 < R $\leq$ 4		\\
Sy1.8 	&	 R > 4				\\
Sy1.9 	&	Broad component only in H$\alpha$ \\
\hline
\end{tabularx}
\end{table}
 
In the second we calculate directly the type through \cite{Netzer}:

\begin{equation} 
\mathrm{Type}\sim1+\bigg[\frac{I(H\alpha_{\mathrm{n}})}{I(H\alpha_{\mathrm{t}})}\bigg]^{0.4}
\end{equation}

\noindent using the properties of the sole H$\alpha$ line. The idea, in both the cases, is to have a comparison between the narrow and the broad/total components of the lines. \\
In the UM framework, IS are positioned between Sy1 and Sy2 galaxies, suggesting they represent AGN with intermediate viewing angles and partial obscuration. More recently, some authors have brought attention to the limitations and challenges of explaining Seyfert galaxy properties solely within the UM framework \cite{Jarvela20}. This study centered on J2118--0732, which is the first confirmed non-local IS residing within an interacting late-type galaxy, capable of maintaining powerful relativistic jets detectable at $\gamma$-ray energies. The probability of observing such a $\gamma$-ray source is exceedingly low within the UM framework. More commonly, relativistic jets are observable in Sy1 due to the alignment of the jet with the observer's line of sight, leading to enhanced emissions through the beaming process. However, a small fraction of AGN, roughly 2.8\% of $\gamma$-ray-detected AGN, exhibit misaligned jets \cite{Foschini22}. The intriguing question that arises is why these misaligned sources constitute such a minor portion of $\gamma$-ray-emitting AGN. Several potential explanations can be considered: these objects may appear as IS, but their line profiles are modified by intrinsic processes unrelated to obscuration, or their scarcity may be due to the faintness of $\gamma$ emissions when they are not aligned with the observer's line of sight, possibly as a result of instrumental limitations. \\ 
From these observations, it becomes evident that the UM alone does not provide a sufficient explanation for the differences among Seyfert classes. One possible approach is the introduction of a hybrid model that combines the extrinsic representation (the UM) with more intrinsic factors such as the presence of relativistic jets, outflows, or evolutionary processes. Jets can significantly alter the spectral characteristics of the source, especially when dealing with Changing-Look AGN (CL-AGN). In the most common scenario, CL-AGN initially exhibit a spectrum dominated by emission lines, which then transforms into a featureless spectrum, and vice versa \cite{Foschini22}. However, this transformation is not the only possibility. Variations in obscuration or accretion rates can also induce changes in the optical spectrum, particularly in the BLR of the emission lines \cite{Risaliti10}, \cite{Storchi}. Alternatively, outflows can reshape the profiles of lines, such as [O III]$\lambda\lambda$4959, 5007. These oxygen lines often display blue bumps or additional components, which are associated to outflows. This can have a significant impact on the classification process, given that [O III]$\lambda$5007 is one of the most critical lines for both manual and automated (machine learning) classification methods \cite{Peruzzi}. Lastly, we can consider an evolutionary scenario in which a peculiar class of Seyferts (Narrow-Line Seyfert 1, NLS1) can be seen as young or rejuvenated quasars that transition into their mature phase and follow an evolution path known as the blazar sequence \cite{Fossati}, \cite{Foschini14}. \\
Another aspect related to the intrinsic hypothesis is the difficulty in classifying Seyfert types accurately. We have previously discussed the most common classification methods, both of which rely on distinguishing between the emission from the BLR and the NLR. However, a significant problem arises when the spectrum has a low Signal-to-Noise ratio (S/N), as the edges of each emission component can overlap, potentially leading to misclassification. This highlights the limitations of the current classification methods and underscores the need for further research to better understand the nature of different source classes. \\
Additionally, there is currently a lack of dedicated research focused on IS and an in-depth analysis of their specific properties compared to the main Seyfert types. Occasionally, IS objects can introduce ambiguity into samples of AGN classified as Seyferts. For example, in a study by Berton et al. (2020)  \cite{Berton20} a previously classified NLS1 exhibited a characteristic IS profile for H$\beta$, featuring a combination of broad and narrow components. \\
This interpretational challenge is closely tied to the quality of the spectra. When S/N is particularly low, a faint BLR emission, typical of Sy1.8/1.9, may mix with the continuum, giving the appearance of a Sy2. Conversely, a faint NLR emission, typical of Sy1.2, can be overwhelmed by the broad component, resulting in a Sy1-like line shape for the Balmer lines. To address this ambiguity, one potential solution is to prioritize the study of high S/N spectra, ensuring more accurate classifications. However, this approach may necessitate excluding fainter sources or observations from small telescopes. As always, a balance must be struck between the completeness of the samples and the precision of our results. \\
This study is centered on the analysis of the optical spectra of the selected sources, while the analysis of X-ray spectra of the same sample will be included in a forthcoming paper. Here are presented only preliminary results on this study, the error estimation is not present, but as a reference we calculate it for the worst spectra in the sample (lower S/N), which is CGCG 164-019. In this case we adopted H$\beta$ as representative for the other lines. The error is calculated producing N=100 spectra summing the original one with a random noise proportional to the one between 5050--5150 \r{A} and performing again the fitting obtaining a distribution for the curve parameters (amplitude, wavelength, width). The errors on the fitting parameters are $\leq$10\%. \\
The paper is organized as follows: in section 2 the {\it{Sample selection}}, in section 3 the {\it{Spectral analysis}}, in section 4 the {\it{BPT/VO diagrams}}, in section 5 the  {\it{Line ratios}}, and in section 6 the {\it{Conclusions}}. \\
In this paper we assume a standard $\Lambda$CDM cosmology with H$_0$=73.3 km s$^{-1}$ Mpc$^{-1}$, $\Omega_{matter}$=0.3, and $\Omega_{vacuum}$=0.7 \cite{Riess}.

\section{Sample selection}
To obtain a reliable sample of Seyfert galaxies, we selected two complementary samples using their BASS or SDSS spectra, so relying on their optical emission properties. The first sample, obtained from the Swift-BAT AGN Spectroscopic Survey (BASS)\footnote{\url{http://www.bass-survey.com}}, comprises hard X-ray-selected sources. The hard X-ray emission primarily originates from the corona of the accretion flow, interacting with cold material through inverse Compton scattering of optical/UV photons on relativistic electrons \cite{Peterson}. At the same time, soft X-ray are absorbed by the neutral hydrogen, hardening the spectra. Consequently, according to the UM, these sources tend to be biased toward high viewing angles, such as almost edge-on orientations. \\
The second group of Seyfert galaxies was selected from the study by Foschini et al. \cite{Foschini22}, which is a $\gamma$-ray selected sample from the 4th Fermi Large Area Telescope catalog (4FGL) \cite{Abdollahi}. The $\gamma$-ray emission is closely related to the only AGN component capable of producing these energetic photons: the relativistic jets. In the UM framework, these objects are biased toward small viewing angles, resembling face-on orientations, with relativistic jets almost aligned with the observer's line of sight. Our study involves the objects from \cite{Foschini22} labeled as: SEY (Seyferts), MIS (misaligned), CL (changing-look), NLS1, and AMB (ambiguous). \\
These sample selections complement each other due to the different selection methods: the first sample prefers edge-on sources, often associated with Sy2 or high-intermediate types (Sy1.8/1.9), while the second favors face-on galaxies that can be identified as Sy1 or low-intermediate types (Sy1.2/1.5).\\
Following the preliminary selection (see the second row in Table~\ref{tab3}), all spectra were meticulously inspected one by one and categorized into partial and complete samples. The first group includes sources where only H$\beta$ and [O III]$\lambda\lambda$4959, 5007 were visible, while the second group encompasses sources with a spectral range from H$\beta$ to [S II]$\lambda\lambda$6717, 6731, which is ideal for constructing the Baldwin, Phillips \& Terlevich/Veilleux \& Osterbrock (BPT/VO) diagrams \cite{BPT}, \cite{VO}. These diagrams can distinguish between starburst objects and AGN, as well as classifying AGN into Seyfert and Low-Ionization Nuclear Emission-line Region galaxies (LINERs) \cite{Heckman}. They rely on the ratio between the Balmer lines and the ionized forbidden lines, revealing the presence of a more or less powerful radiation field, which, in turn, is related to the presence or absence of a central emitting region, the AGN. \\
As an initial approach, we considered only the complete spectra, enabling us to construct the BPT/VO diagrams. At this point, the two samples consisted of 121 sources from the BASS selection and 50 objects from the 4FGL selection. \\
We excluded Sy1.9 sources from the sample due to the Balmer decrement calculation, which relies on the ratio between the broad components of both H$\alpha$ and H$\beta$, a feature absent in Sy1.9 (only present in H$\alpha$). Consequently, we obtained 38 AGN from BASS and 7 sources from 4FGL. A summary of these steps is provided in Table~\ref{tab3}. \\
The limited number of sources in the 4FGL sample results from our reliance on publicly available spectra from the Sloan Digital Sky Survey (SDSS). We supplemented our dataset with two additional objects from the 4FGL sample, namely 1H 0323+342 and PKS 2004-447, both NLS1 sources. These spectra come from the Boller \& Chivens spectrograph mounted on the Galileo telescope in Asiago (University of Padova) and from the FOcal Reducer/low dispersion Spectrograph 2 (FORS2) at UT1 (Very Large Telescope, European Southern Observatory), respectively. They were added to provide a reference for the NLS1 group in terms of obscuration, Seyfert type, and disk luminosity. Both of them respect the same conditions for the spectra selection, even if 1H 0323+342 show very faint [S II]$\lambda\lambda$6717, 6731 lines.

\begin{table}[h!] 
\caption{Selection steps operated to obtain the final sample using the BASS and SDSS optical spectra of the sources.\label{tab3}}
\begin{tabularx}{\textwidth}{cc|cc}
\hline
\multicolumn{2}{c|}{\textbf{4FGL}} & \multicolumn{2}{c}{\textbf{BASS}}\\
\hline
\multicolumn{1}{c}{\textbf{$\quad$Number}} & \textbf{Condition} & \textbf{$\quad$Number} & \multicolumn{1}{c}{\textbf{Condition}} \\
\hline
  &	 Total number of sources 	&	& \\
$\quad$2982 & from 4FGL (with  & 1210& Total number of sources\\
& 1H 0323+342 and & &  from BASS\\
&  PKS 2004-447) & & \\
& & & \\
$\quad$80	&	 CL + MIS + SEY	&	638	&	With the optical spectra\\
& + NLS1 + AMB & & \\
 & & & \\
 &	 With at least some	&	&	With at least some\\
$\quad$52 & spectral features visible & 334 & spectral features visible\\
& (H$\alpha$/H$\beta$/[O III]) & & (H$\alpha$/H$\beta$/[O III]) \\
 & & & \\
$\quad$15 	&	 Complete spectra	&	121	&	Complete spectra \\
& (from H$\beta$ to [S II]) & & (from H$\beta$ to [S II]) \\
 & & & \\
 &	 No Sy1.9, [O I]$\lambda$6300 	&	&	 No Sy1.9, [O I]$\lambda$6300 \\
$\quad$9 & and/or [S II]$\lambda\lambda$6717, & 38 & and/or [S II]$\lambda\lambda$6717, \\
& 6731 significant & & 6731 significant \\
\hline
\end{tabularx}
\end{table}

\begin{figure}[h!]
\includegraphics[width=12 cm]{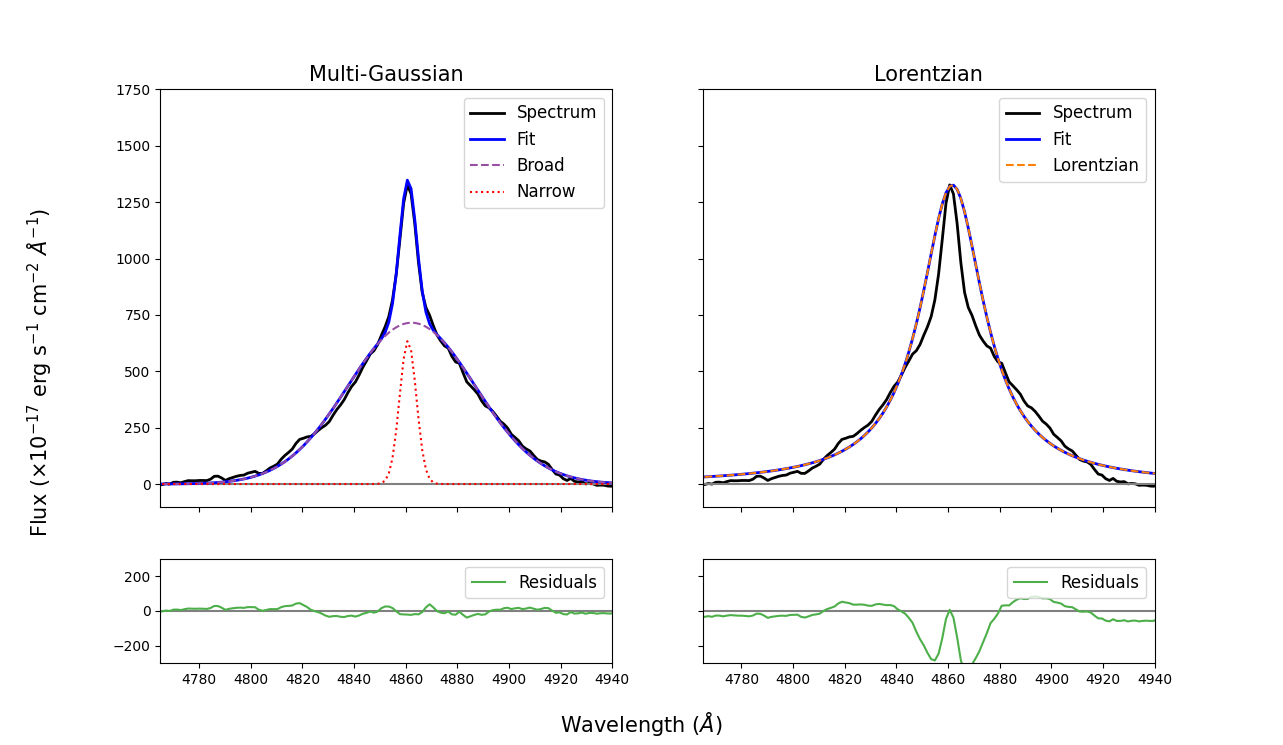}
\centering
\caption{Comparison of the H$\beta$ fitting for Mrk 79. In the left panel, a multi-Gaussian approach is used with a curve for the BLR emission (dashed-violet) and a second curve for the NLR one (dotted-red). The sum of the two components is shown in blue, and in the lower panels, the residuals in green represent the difference between the adopted model and the spectra. In the right panel, the same analysis is conducted with a Lorentzian. It is evident that in the second case, the residuals are more pronounced and tend to overestimate both the center and the wings of the line. \label{fig0}}
\end{figure}  

\begin{figure}[h!]
\includegraphics[width=10 cm]{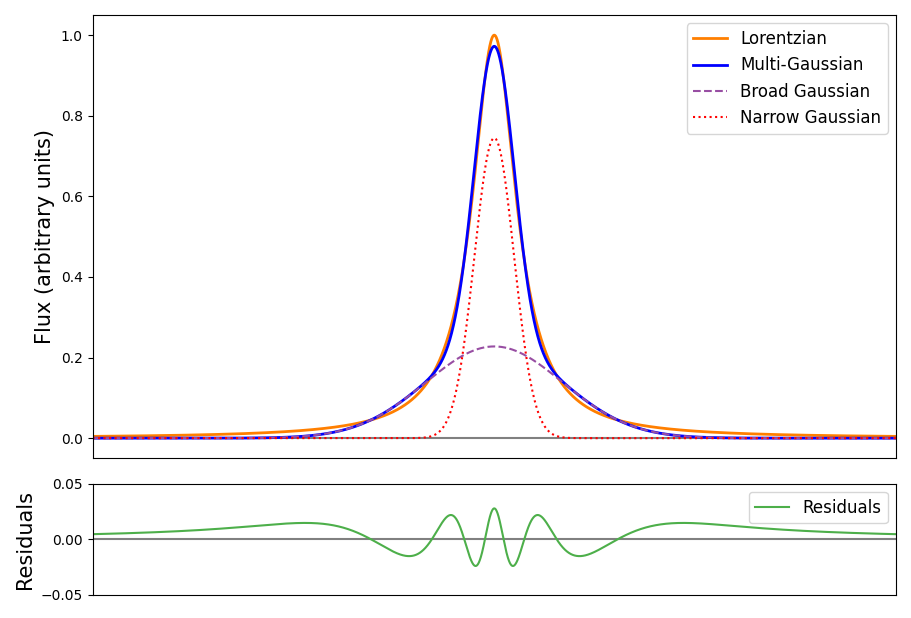}
\centering
\caption{Comparison between a Lorentzian profile (in orange) and a multi-Gaussian one (in blue, with the narrow and broad components shown in red/violet) with a FWHM of approximately 2000 km s$^{-1}$. The lower panel illustrates the residuals as the difference between the double Gaussian model and the Lorentzian profile. \label{fig1}}
\end{figure}  
\begin{figure}[h!]
\includegraphics[width=14 cm]{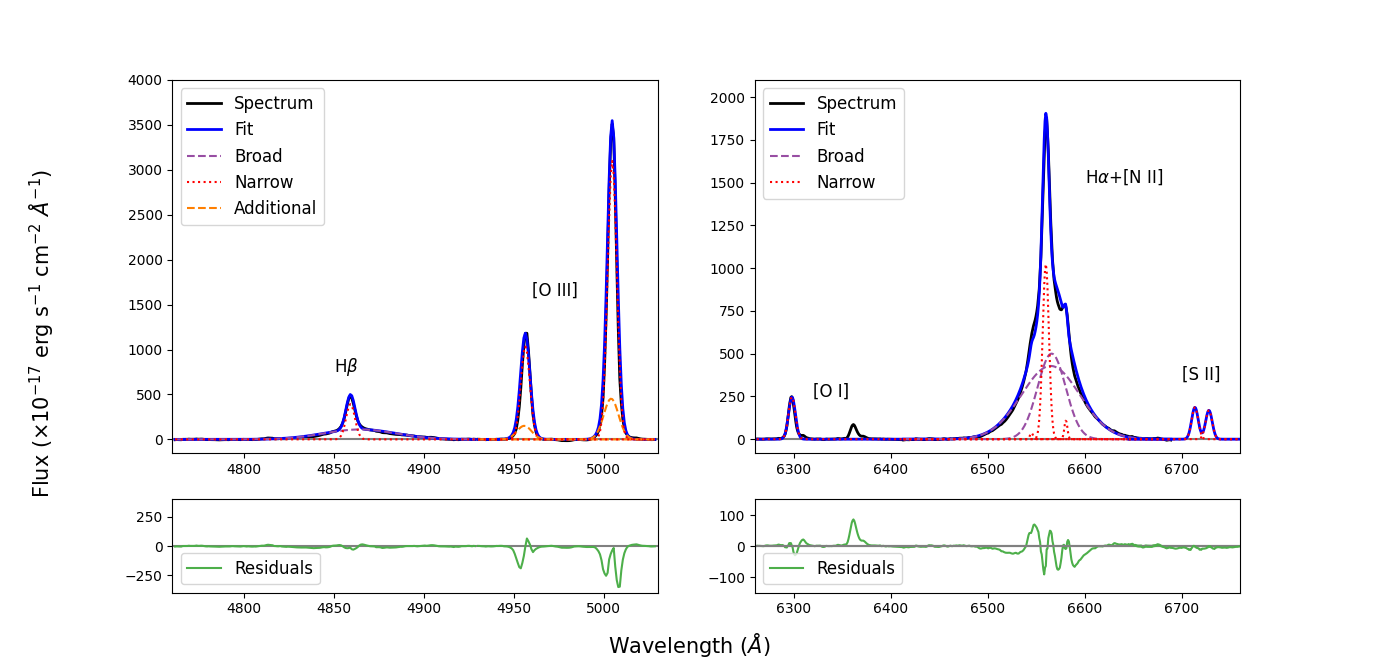}
\centering
\caption{From left to right: fitting of H$\beta$, [O III]$\lambda\lambda$4959, 5007, [O I]$\lambda$6300, H$\alpha$, [N II]$\lambda\lambda$6548, 6583, [S II]$\lambda\lambda$6717, 6731 for Mrk 110. As in Figure~\ref{fig1}, the blue line represents the total fitting, the red dotted curves represent the narrow component, the violet dashed curves represent the broad component(s), and the orange dashed curve is associated with a possible outflow component related to the [O III] emission. These blueshifted bumps are commonly observed in the [O III] profiles \cite{Boroson}. In the lower panel, the residuals are shown in light green. In the bottom-right panel of the figure, the [O I]$\lambda$6363 line is present as a residual because it was not fitted during the analysis. \label{fig2}}
\end{figure}  

\begin{figure}[h!]
\includegraphics[width=14 cm]{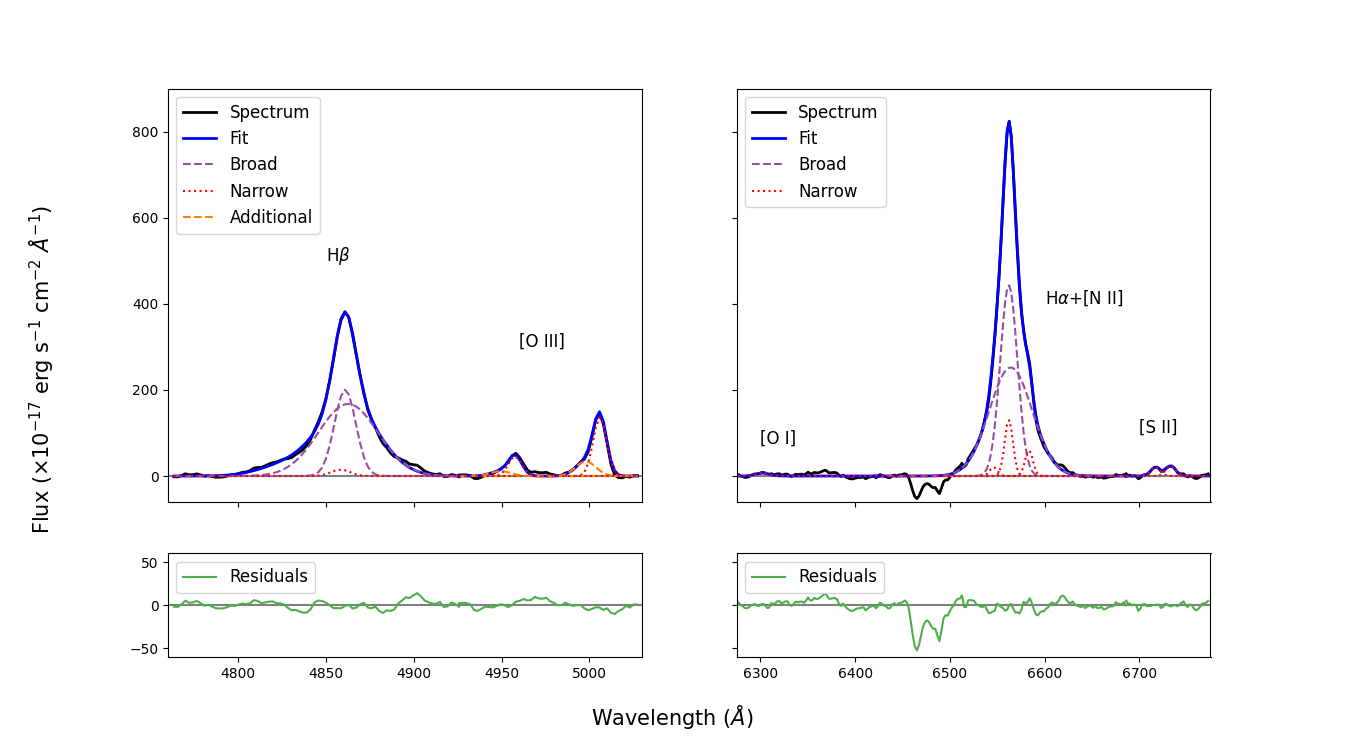}
\centering
\caption{Fitting of the spectrum of 1H 0323+342 in the H$\beta$--[O III]$\lambda\lambda$4959, 5007 region is shown in the left panel, and of the [O I]$\lambda$6300--[S II]$\lambda\lambda$6717, 6731 region for 1H 0323+432. The line styles and colors are the same as in Figure~\ref{fig2}. Additionally, there is a telluric absorption feature between 6450-6500 \r{A}. \label{fig3}}
\end{figure}  

\section{Spectral analysis}
\subsection{Preliminary steps}
The spectral analysis followed the classical initial steps, including cosmological redshift correction, de-reddening, host galaxy subtraction, and iron line subtraction. Each of these steps was implemented using Python 3.0 scripts, and the specific tools used in each step are detailed below. \\
For the cosmological redshift correction, we utilized a built-in function available on GitHub\footnote{\url{https://github.com/sczesla/PyAstronomy/blob/master/src/pyasl/asl/dopplerShift.py}}. \\
The de-reddening procedure was based on the Cardelli, Clayton, Mathis (CCM) extinction law \cite{CCM} and the relationship between $A(V)$ (total extinction in magnitudes) and $N_H$ (column density of neutral hydrogen atoms in cm$^{-2}$) \cite{Bohlin}. To calculate $N_H$, we used the dedicated NASA-HEASARC tool\footnote{\url{https://heasarc.gsfc.nasa.gov/cgi-bin/Tools/w3nh/w3nh.pl}}, which provides a weighted value. We then converted this value into $A(V)$ using the proportionality factor of $5.3 \times 10^{22}$ mag cm$^{-2}$ mentioned earlier and applied it within the CCM extinction law. Finally, we used:

\begin{equation}
	I_{\lambda,0}=I_{\lambda}\cdot10^{0.4\cdot A(\lambda)}
\end{equation}
 
The host galaxy subtraction was carried out exclusively for sources with a  redshift $\lesssim$ 0.3 \cite{Letawe}, and it was applied to spectra where the absorption features were clearly distinguishable. We employed galaxy templates from Mannucci et al. \cite{Mannucci}, which encompass various host galaxy types, including elliptical and spiral galaxies, with a spectral resolution of 5 \r{A}. Due to the template's inherent wavelength binning, we needed to rebin the AGN+host spectra before conducting the subtraction. In general, we used the Ca II lines at 3934 \r{A} and 3968 \r{A} as a reference to renormalize the template. These lines are exclusively produced by the galaxy's stellar components. \\
For the iron subtraction, we made use of templates from Kovacevic et al. (2010) and Shapovalova et al. (2012) \cite{Kovacevic}, \cite{Shapo}, which are accessible on the Serbian Virtual Observatory\footnote{\url{http://servo.aob.rs/FeII\_AGN/link6.html}}. These templates span the wavelength range from 4000 \r{A} to 5500 \r{A} and include various Doppler widths of Fe II lines, ranging from 700 to 2800 km s$^{-1}$. We performed iron subtraction in spectra where the peaks at 4570 \r{A} and 5270 \r{A} exceeded the continuum value by at least 2$\sigma$ (standard deviation).

\subsection{Line fitting}
Similar to the iron subtraction, line fitting was performed with a detection limit set at >2$\sigma$. This constraint contributed to the limitation in the number of sources analyzed. This procedure proved especially useful in cases where the considered line was very faint and closely intertwined with noise contributions to the flux. \\
To construct the BPT/VO diagram, we employed a multi-Gaussian model for fitting the following lines: H$\beta$, [O III]$\lambda\lambda$4959, 5007, [O I]$\lambda$6300, [N II]$\lambda\lambda$6548, 6583, H$\alpha$, and [S II]$\lambda\lambda$6717, 6731. We chose the multi-Gaussian approach due to the composite nature of the spectra, as previously mentioned, which includes both a broad and a narrow emission component (BLR and NLR, respectively). The narrow peaks of H$\beta$ and H$\alpha$ occur in the same region as the forbidden lines (e.g., [O III]). Consequently, the Full Width at Half Maximum (FWHM) of these lines remains fixed, limiting the number of free parameters. An alternative to Gaussian fitting is the Lorentzian approach, but it does not adequately fit the composite profile of these spectra. Specifically, it can overestimate the contribution of the wings when the narrow component dominates or overestimate the peak when the broad component is prominent. A comparison between the multi-Gaussian and Lorentzian approaches is illustrated in Figure~\ref{fig0}. However, selecting the correct model is not always straightforward, as seen in Figure~\ref{fig1}, where a double Gaussian and a Lorentzian yield similar profiles when the FWHM is approximately 2000 km s$^{-1}$. This is particularly relevant because 2000 km s$^{-1}$ marks the boundary between classifying an object as a NLS1 or a Sy1. The ability to distinguish between these two classes is highly dependent on the resolution of the analyzed spectra, as emphasized by Berton et al. \cite{Berton20}. \\
To minimize the number of free parameters in the fitting process, we imposed some constraints. For instance, starting with the [O III]$\lambda\lambda$4959, 5007 lines, we allowed the most prominent line ([O III]$\lambda$5007) to be fitted freely while binding all parameters (center, FWHM, and amplitude) of the other line. The center displacement and FWHM were the same for both lines, and the amplitude of [O III]$\lambda$4959 was fixed at one-third of the amplitude of [O III]$\lambda$5007 \cite{Dimitri}. Similar FWHM constraints were applied between [O III] and the narrow component of H$\beta$ and H$\alpha$. When dealing with the H$\alpha$+[N II] complex, constraints were applied to the amplitudes of the two nitrogen lines and the FWHM. The amplitude ratio between [N II]$\lambda$6548 and [N II]$\lambda$6583 was set to 1/2.96 \cite{Tachiev} (NIST database\footnote{\url{https://www.nist.gov/pml/atomic-spectra-database}}), and the FWHM was kept consistent with the previous cases. The same procedure was employed for the [S II]$\lambda\lambda$6717, 6731 lines, where the FWHM was identical for both lines. An example of the line fitting using this procedure is depicted in Figure~\ref{fig2}. The fitting process was conducted using a dedicated Python code based on the Levenberg-Marquardt algorithm\footnote{\url{https://docs.astropy.org/en/stable/modeling/}} and least squares statistics.

\subsection{Notes on the sample}
\begin{figure}[h!]
\includegraphics[width=14 cm]{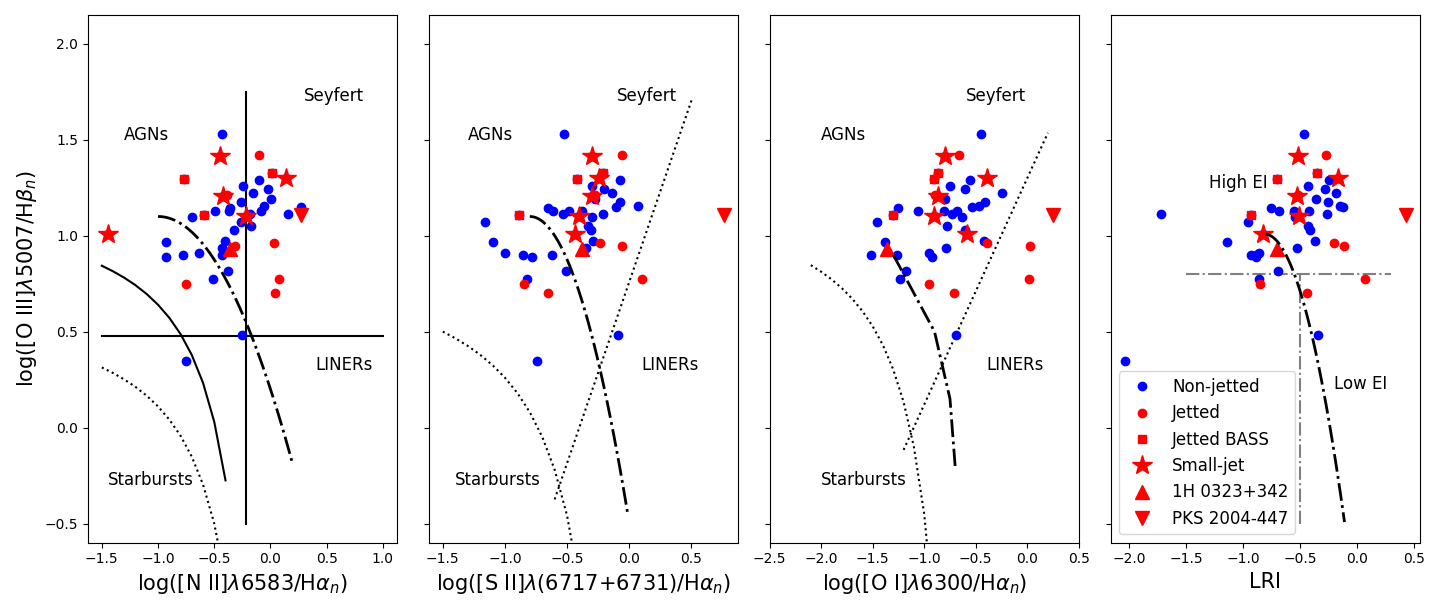}
\centering
\caption{The first three plots are the BPT/VO diagrams for the selected sources while the last one is from Buttiglione et al. \cite{Buttiglione} and uses the LRI. In blue the non-jetted objects (BASS) and in red the jetted ones (mainly 4FGL, some BASS). The jetted sources are marked with different symbols according to their properties: the red dots are the classical jetted AGN, the red squares are from BASS, the red stars present small jets in the parsec scale, and with the triangles we have 1H 0323+432 and PKS 2004-447. The solid lines identify the classification limits from Kauffmann et al. \cite{Kauffmann}, the dotted lines from Kewley et al. \cite{Kewley}, and the dashed-dotted lines (both black and grey ones) from Buttiglione et al. \cite{Buttiglione}.\label{fig4}}
\end{figure}

\begin{figure}[h!]
\includegraphics[width=14 cm]{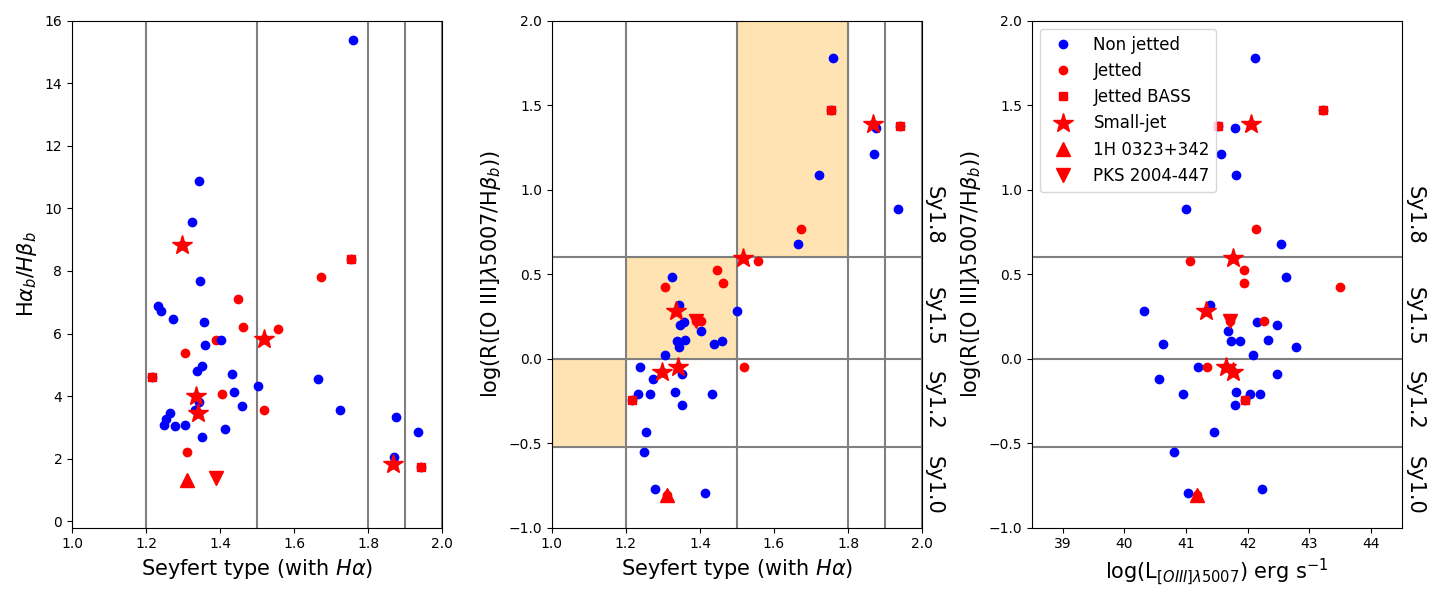}
\centering
\caption{Line ratios for our sample, the markers are the same as in Figure~\ref{fig4}. The colored areas represent the correspondence between the two Seyfert classification methods adopted (from [O III]/H$\beta$ and H$\alpha$ components). The grey solid lines show the separations between the Seyfert classes. The last two plots in the right part share the same y-axis and identify the same Seyfert sub-classes indicated between the diagrams.\label{fig5}}
\end{figure}

As stated at the beginning of \textit{Section 3.2}, some spectra were not considered in the sample due to the faintness of the lines, in particular [O I]$\lambda$6300 and [S II]$\lambda\lambda$6717, 6731. The 2$\sigma$ criterion, together with the necessity of using the same sources in all the plots, further reduced the number of objects. In the end, we obtained: 7 AGN from the 4FGL sample with SDSS (jetted, marked with red dots in Figures~\ref{fig4} and ~\ref{fig5}) and 38 sources from the BASS one (non-jetted, marked with blue dots). We also added two jetted NLS1 from 4FGL with other optical spectra: 1H 0323+342 and PKS 2004-447 (marked with red triangles). From the BASS sample, we noticed 3 jetted AGN: 4C +29.30, 3C 227, and 3C 234.0 (again marked with red dots); and 5 objects with parsec-scale jets (small-jet, marked with red stars): 2MASX J04234080+0408017 \cite{Smith}, Mrk 590 \cite{Yang}, NGC 985 \cite{Doi}, Mrk 110 \cite{Jarvela22}, and Mrk 705 \cite{Jarvela22}. It is important to notice that there is no superposition between the two samples. \\
For the spectra of 1H 0323+342, it is important to stress that this source was monitored by one of us between 2014 and 2021 with the Asiago Galileo telescope, and so a lot of spectra were available. To increase the S/N, we combined some of them, applied the same steps listed before (de-reddening, cosmological redshift, and iron subtraction), and performed the line fitting. The result is shown in Figure~\ref{fig3}: both [O I]$\lambda$6300 and [S II]$\lambda\lambda$6717, 6731 are very faint.
As previously mentioned, these are preliminary results, and in the near future, one of the further steps will be including more jetted sources due to their limited number in this analysis. 

\section{BPT/VO diagrams}
The BPT/VO diagrams are employed as a tool for the separation between AGN and starburst galaxies using the properties of different ionization lines. Many authors have proposed slightly different classifications \cite{Kauffmann}, \cite{Kewley}, but they always identify three regions: starbursts, Seyferts, and LINERs. These separations can be seen in Figure~\ref{fig4} with the solid and dotted lines. The dashed-dotted lines represent the limits for the sources in Buttiglione et al. \cite{Buttiglione}, which are spread in the upper-right part of the curve. In their paper, they studied the optical properties of a homogeneous sample of the 3CR radio sources with a redshift < 0.3. Consequently, this population characterizes the mean characteristics of jetted sources. Of course, this dashed-dotted line serves as a rough reference used to compare a pure jetted sample with ours. \\
The last plot in Figure~\ref{fig4}, the line ratio index (LRI) vs. [O III]/H$\beta$, is also sourced from \cite{Buttiglione} and represents a composite view of the three BPT/VO diagrams. The y-axis aligns with the diagnostic diagrams, while the x-axis is calculated as follows:

\begin{equation}
	\mathrm{LRI}=\frac{1}{3}\bigg(\log\frac{\mathrm{[N II]}}{\mathrm{H}\alpha}+\log\frac{\mathrm{[S II]}}{\mathrm{H}\alpha}+\log\frac{\mathrm{[O I]}}{\mathrm{H}\alpha}\bigg)
\end{equation}

\noindent The objective here is to create a more stable index than single-line ratios, aimed at avoiding ambiguous classifications, as emphasized by Kewley et al. \cite{Kewley}. In the LRI-$\log$([O III]/H$\beta$) plot, two distinct regions are separated by a horizontal line at approximately 1: the high Excitation Index (EI) and low EI regions. EI represents the overall ratio of high and low excitation emission lines in each source. While these two populations are clearly visible in Buttiglione et al. \cite{Buttiglione}, our work primarily focuses on sources concentrated in the high ionization region. The absence of low ionization sources may be related to the nature of the selected sources (AGN) and the methods used to obtain the sample, based on $\gamma$ and hard-X emissions. \\
In Figure~\ref{fig4}, it is evident that the majority of the sample falls within the Seyfert region, with some exceptions in the LINERs region. Particularly, objects in the LINERs area can be interpreted as AGN undergoing a transition from an efficient to an inefficient phase in the accretion process \cite{Kewley}, possibly nearing the end of an activity phase. \\
We also explored the possibility of replacing line fluxes with peak values to address modeling issues, but this did not prove to be a valid alternative. For further details, please refer to Appendix B.

\section{Line ratios}
The objective is to investigate the relationship between the intrinsic properties of AGN and obscuration. The oxygen luminosity (L$_{\mathrm{[O III]}}$) can be interpreted as a proxy for the disk luminosity (L$_{\mathrm{disk}}$) \cite{Heckmann}, \cite{Risaliti}, \cite{Marin}. Some authors criticized this statement \cite{Haas}, \cite{Baum} demonstrating that in some cases also [O III] is partially obscured in type 2 AGN, so it is not totally independent from the classification type of the source. In the optical range the [O III] luminosity is the sole quantity that we can use as a tool for the disk one. Expanding the analysis to the X-rays we can eliminate this problem producing more accurate conclusions. With this aim, we are analyzing also the X-ray part of the spectra of the same objects to obtain more solid informations about the possible connection between the disk luminosity and the classification (B. Dalla Barba et al., in prep). \\
To assess obscuration we utilize the Balmer decrement and Seyfert type as tools. The Balmer decrement is frequently employed as a means to assess the degree of obscuration because its intrinsic value remains relatively insensitive to variations in gas temperature and density within the region being examined. However, the total value, which is what we observe, includes the effects of obscuration. For the case B recombination in HII regions, the reference value for H$\alpha$/H$\beta$ is approximately 2.9 \cite{Osterbrock89}. In contrast, for AGN, this theoretical value can vary depending on the Seyfert type, as emphasized by various authors \cite{Binette}, \cite{Nagao}, \cite{Schnorr}. Consequently, if we wish to disentangle the level of obscuration from the overall H$\alpha$/H$\beta$ ratio, we require a reliable estimate of the intrinsic value. This task is particularly challenging when the sample comprises different Seyfert types. Focusing the study on unobscured AGN, such as Sy1, some researchers \cite{Dong}, \cite{LaMura} have discovered a general consensus between the Balmer decrement values and the case B recombination. The values typically fall within the 3.0-3.5 range with a dispersion of 0.2-0.6. There are, however, some outliers with decrements of 1.6 or 7.8. Consequently, we will consider the case B recombination as a valid reference, though it is important to bear in mind that the conclusions regarding the Balmer decrement should account for the Seyfert classification. Furthermore, it is worth noting that the intrinsic value may be higher than the commonly used value of 2.9. \\
The link between L$_{\mathrm{[O III]}}$ and L$_{\mathrm{disk}}$ arises from the level of activity of the central engine and the degree of ionization produced in the NLR. Additionally, the stellar component of the host galaxy can contribute to the oxygen flux, particularly when the orientations of the host and AGN are not aligned \cite{Schmitt}. This different inclination between the host and the AGN was studied by Kinney et al. and Nagar \& Wilson concluding that the difference in viewing angle to the central engine (using the radio jet orientation) between Sy1 and Sy2 is a direct and independent confirmation of the UM \cite{Kinney}, \cite{Nagar}. Some explanations for this misalignment can be the warping of the accretion disk by self-irradiation instabilities or the presence of a second gravitating system as during a merger. This does not apply to our case because the systems in question are not experiencing significant instabilities, except for UGC 8327 which is a dual system. \\
As previously mentioned, the inclusion of X-ray spectral analysis will provide an independent measure of obscuration. This method will help us disentangle the influence of Seyfert types on the Balmer decrement and address any host galaxy-AGN misalignment issues previously discussed.\\
For the moment, we have generated three plots in Figure~\ref{fig5}, which will be discussed in the following subsections.

\subsection{Seyfert type vs Balmer decrement}
The initial approach was to compare the Seyfert type with the BLR Balmer decrement (H$\alpha$/H$\beta$). H$\alpha$/H$\beta$ provides an estimate of the extinction level because the decrement's value is closely related to gas physical properties. Consequently, discrepancies from the theoretical value can be attributed to obscuration effects, even if some authors argue this choice \cite{Binette}, \cite{Nagao}, \cite{Schnorr}. In the first panel of Figure~\ref{fig5}, the results are displayed: the non-jetted and jetted populations do not exhibit clear separation. This suggests that there are no significant differences in terms of the Balmer decrement and classification. This result poses an interpretative challenge. \\
The Balmer decrement is commonly used as an indicator of obscuration, with Sy1/ Sy1.2/Sy1.5 expected to be less obscured (lower Balmer decrements) than Sy2/ Sy1.9/Sy1.8 (higher Balmer decrements). If the separation between jetted/non-jetted follows the UM scheme and, consequently, corresponds to face-on/edge-on sources (e.g., Sy1/Sy2), the Balmer decrement should display an ascending trend in the second plot of Figure~\ref{fig5}, which is not observed. The results for the statistical tests are as follows: Spearman (r$_p$=-0.10, p-value=0.46), Pearson (r$_p$=-0.07, p-value=0.61), and Kendall (r$_k$=-0.06, p-value=0.53). In this case, there is no apparent correlation. \\
One possible explanation for the lack of a correlation between the Balmer decrement and the Seyfert type can be the dependence of H$\alpha$/H$\beta$ with the radiation field properties. In this context we can have an excess of H$\alpha$ emission relative to H$\beta$ produced by an intense source of high-energy radiation (the AGN) \cite{Gaskell}. \\
Another point to stress is the possible bias introduced by the lack of Sy1.9 sources in the sample, for these objects it is not possible to calculate H$\alpha_b$/H$\beta_b$ due to the missing broad component in H$\alpha$. From the first panel of Figure~\ref{fig5} we can see some objects in the Sy1.9 region, this is due to the classification method adopted. As we will explain in section 5.2, [O III]/H$\beta$ trace better the Seyfert type than H$\alpha$. \\
Finally, also the influence of the Seyfert type on the intrinsic value of the Balmer decrement can play a role as stated in the previous section. \\

\subsection{Seyfert type vs [O III]/H$\beta$}
A second approach involved comparing different Seyfert classification methods. In the second panel of Figure~\ref{fig5}, it is evident that the two methods generally agree. For visual clarity, we identify the colored areas as regions where the two classification methods coincide. Almost all the data points are distributed around the diagonal, differing by at most one Seyfert class along both axes. In particular, objects categorized as Sy1/1.2 on the y-axis exhibit a tendency to be "over-classified" on the x-axis, indicating higher Seyfert types. For example, 1H 0323+342, a widely recognized NLS1, is classified as a Sy1 based on [O III]/H$\beta$ measurements, but as a Sy1.2/1.5 when considering H$\alpha$ properties. While the difference is not extreme, it is notable. This implies that the H$\alpha$-based classification method may not be as reliable as the [O III]--H$\beta$ approach, especially for lower Seyfert types. The issue with the H$\alpha$ method may stem from challenges in accurately fitting the H$\alpha$+[N II] complex or from intrinsic characteristics of the BLR. \\
Statistical tests revealed a strong trend: Spearman (r$_p$=0.77, p-value=3.57$\cdot 10^{-10}$), Pearson (r$_p$=0.85, p-value=5.77 $\cdot10^{-14}$), and Kendall (r$_k$=0.59, p-value=6.44 $\cdot10^{-9}$). One important thing to stress is the absence of separation on the horizontal axis between jetted and non-jetted AGN. Our sample is based on 4FGL and BASS objects, so the jetted sources are selected using high-energy part of the spectrum. With this in mind, according to the UM, this type of jetted sources should be biased towards lower IS types (such as Sy1.2/1.5), and non-jetted ones towards higher types (such as Sy1.8/1.9). However, this is not visible in this plot.\\
Globally, [O III]/H$\beta$ show a trend with the Seyfert type (with H$\alpha$), see the second panel in Figure~\ref{fig5}. In the UM framework, an increasing Seyfert type is interpreted as indicating an higher level of obscuration (from Sy1 to Sy2). Consequently, we will adopt [O III]/H$\beta$ as a proxy for the obscuration level. The same correspondence is not observed for the Balmer decrement.

\subsection{L$_{\mathrm{[OIII]}}$ vs [O III]/H$\beta$}
The final relationship we investigated is between oxygen luminosity (L$_{\mathrm{[O III]}}$) and [O III]$\lambda$5007/H$\beta$. As mentioned earlier, L$_{\mathrm{[O III]}}$ serves as a tool for intrinsic properties, particularly disk luminosity, while the obscuration level is better estimated by [O III]/H$\beta$. The pros and cons of these quantities were discussed in section 5, here we assume that both of them are good tracers of the intrinsic and obscuration properties respectively of the selected AGN. \\
The results from the last panel of Figure~\ref{fig5} reveal that these two properties are not related. There is no significant correlation observed in this comparison and the two populations are mixed. This suggests that Seyfert type (and subtype) does not depend on L$_{\mathrm{[O III]}}$, and consequently, the intrinsic properties that can be derived from the optical range of the spectrum. In other words, this implies that the properties of the central region of the AGN do not lead to differences in terms of Seyfert type, supporting the idea of a common structure, as predicted by the UM. The results of the statistical tests are as follows: Spearman (r$_p$=0.22, p-value=0.13), Pearson (r$_p$=0.26, p-value=0.08), and Kendall (r$_k$=0.15, p-value=0.15).

\section{Conclusions and future work}
We conducted a search for a possible relationship between the intrinsic properties and the obscuration of a sample of Seyfert galaxies, categorized as jetted and non-jetted sources. Confirming their classification as AGN, particularly as Seyfert for the majority, was achieved through the BPT/VO diagram. Objects situated within the LINERs region can be interpreted as aging AGN in their final activity phase. Additionally, we created the LRI-[O III]/H$\beta$ plot, where nearly all the data points fall within the high EI region. This trend likely stems from the nature of the sources (AGN) and the spectral bands used for sample selection.\\
Next, we generated line ratio plots with the initial aim of finding a suitable parameter related to obscuration level. Our first attempt, using the Balmer decrement, did not yield significant results. However, in a subsequent analysis, we employed [O III]/H$\beta$, which provided valuable insights when compared to Seyfert type, as expected from the UM. As for intrinsic properties, optical spectroscopy limits us to using L$_{\mathrm{[O III]}}$, which can be an indicator of the disk luminosity and, in turn, depends on the central engine's activity. A comparison of these two parameters did not reveal a significant relationship, suggesting that the central engine's structure in Seyfert galaxies may be consistent across all types, with obscuration remaining the most plausible explanation. Summarizing, we do not find any clear separation between the jetted/non-jetted sources in terms of the considered parameters, this finding aligns with the UM. \\
On the contrary, there is no clear separation between jetted and non-jetted objects in terms of IS sub-type. In panel b of Figure~\ref{fig5}, a clear distinction between Sy1.2/1.5 and Sy1.8/1.9, based on $\gamma$-ray jetted/non-jetted categorization, is not evident. This implies that jetted sources do not always correspond to lower Seyfert sub-types (such as Sy1.2 and Sy1.5) but can sometimes correspond to Sy1.8/1.9. This unpredicted result challenges the UM and warrants further investigation. \\
One potential future step involves incorporating X-ray analysis of the same sources to calculate the parameter N$_\mathrm{H}$. Combining optical and X-ray analyses allows for the use of the low-energy part as an estimator of obscuration (using [O III]/H$\beta$ or the Balmer decrement) and the high-energy component as a more robust indicator of accretion power. \\
Another approach is to focus the analysis on specific sources, utilizing integral field spectra. Datacubes enable the study of each region of the AGN, providing both spectra and photometry simultaneously.\\
\\
\noindent {\small{
\textit{Acknowledgments:} BDB is grateful to the organizers of the 14th Serbian Conference on Spectral Line Shapes in Astrophysics (14SCSLSA) for their partial economic support for the conference. This work has been partially funded by INAF-OA Brera Basic Research. Funding for the Sloan Digital Sky Survey V has been provided by the Alfred P. Sloan Foundation, the Heising-Simons Foundation, the National Science Foundation, and the Participating Institutions. SDSS acknowledges support and resources from the Center for High-Performance Computing at the University of Utah. The SDSS web site is \url{www.sdss.org}. SDSS is managed by the Astrophysical Research Consortium for the Participating Institutions of the SDSS Collaboration, including the Carnegie Institution for Science, Chilean National Time Allocation Committee (CNTAC) ratified researchers, the Gotham Participation Group, Harvard University, Heidelberg University, The Johns Hopkins University, L’Ecole polytechnique fédérale de Lausanne (EPFL), Leibniz-Institut für Astrophysik Potsdam (AIP), Max-Planck-Institut für Astronomie (MPIA Heidelberg), Max-Planck-Institut für Extraterrestrische Physik (MPE), Nanjing University, National Astronomical Observatories of China (NAOC), New Mexico State University, The Ohio State University, Pennsylvania State University, Smithsonian Astrophysical Observatory, Space Telescope Science Institute (STScI), the Stellar Astrophysics Participation Group, Universidad Nacional Autónoma de México, University of Arizona, University of Colorado Boulder, University of Illinois at Urbana-Champaign, University of Toronto, University of Utah, University of Virginia, Yale University, and Yunnan University.}}

\appendix
\section[\appendixname~\thesection]{Appendix}
\begin{figure}[h!]
\includegraphics[width=14 cm]{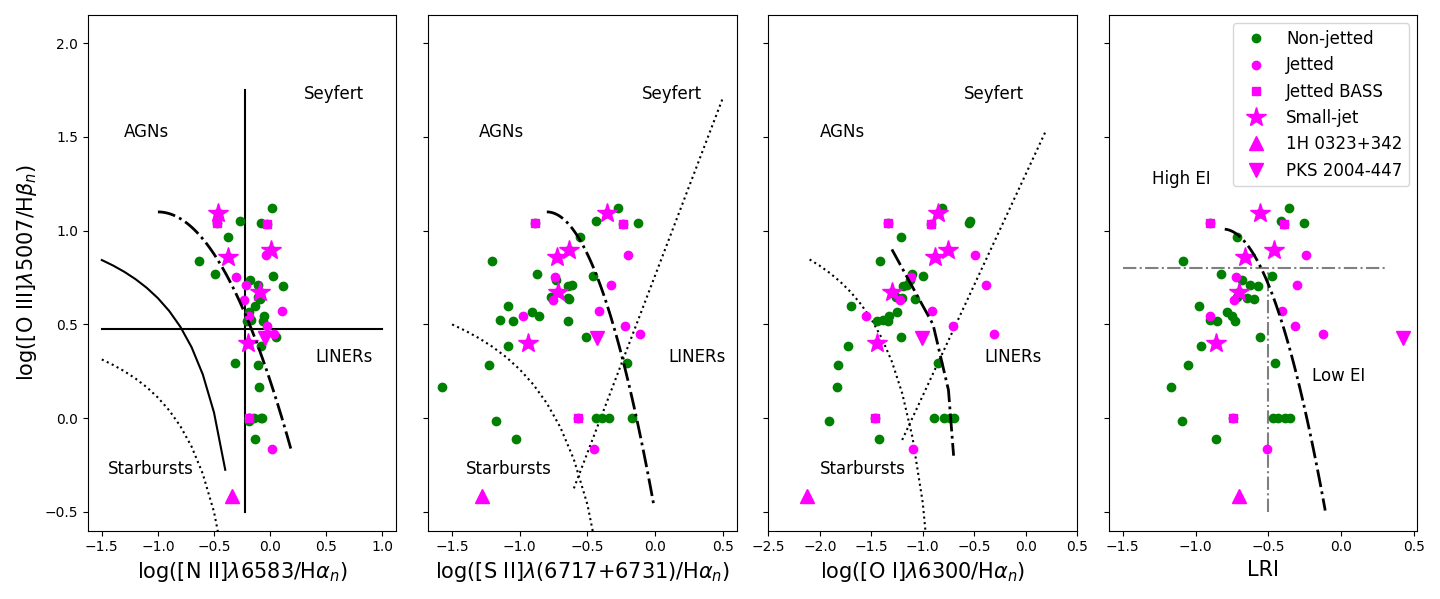}
\centering
\caption{BPT/VO diagram of the selected sample, utilizing the line peak instead of the flux. The line styles and symbols remain consistent with those in Figure~\ref{fig1}, but different colors are employed to emphasize the distinct method employed here. The green points correspond to the blue ones (non-jetted), and the magenta points correspond to the red ones (jetted, small-jets, and NLS1) in the previous BPT/VO diagram (Figure~\ref{fig4}).\label{figa3}}
\end{figure}

Here, we present the results for the BPT/VO diagrams calculated using the value of the peak of the lines, instead of the flux. This idea arises from the potential to avoid the fitting procedure when the spectrum has a low S/N by estimating only the value of the maximum flux for each line. The results are depicted in Figure~\ref{figa3}. When comparing these results with those in Figure~\ref{fig4}, a significant difference is evident between the two approaches. To assess this, we calculated the average shifting of the points in both the vertical and horizontal axes to determine if there is a common trend and if applying a correcting factor is warranted. The average shift (in both axes) and the standard deviation are of a similar order of magnitude, as indicated in Table~\ref{taba1}. Consequently, we can conclude that there is no preferential shift in the positions between the classical BPT/VO and the line-peak BPT/VO plots. The same observation applies to the LRI-$\log$([O III]/H$\beta$) plot. This underscores the fundamental role of the fitting procedure in producing a valid classification of the sources.

\begin{table}[h!] 
\caption{Results for the horizontal and vertical shift (with standard deviation) in the BPT/VO plots as difference between the peak case (Figure~\ref{figa3}) and the flux one (Figure~\ref{fig4}).\label{taba1}}
\begin{tabularx}{\textwidth}{c|cc|cc}
\hline
\multicolumn{1}{c|}{\textbf{ }} & \multicolumn{2}{c|}{\textbf{4FGL}} & \multicolumn{2}{c}{\textbf{BASS}}\\
\hline
\multicolumn{1}{c|}{\textbf{Axis}} & \textbf{Shift} & \textbf{$\sigma$} & \textbf{Shift}  & \multicolumn{1}{c}{\textbf{$\sigma$}} \\
\hline
$ \quad\quad\quad\log$([N II]/H$\alpha$) $\quad$ &	 $\quad$ 0.24 $\quad$ & $\quad$ 0.26 $\quad$ & $\quad$ 0.24 $\quad$ & $\quad$ 0.23 $\quad$\\
& & & & \\
$ \quad\quad\quad\log$([S II]/H$\alpha$) $\quad$ &	$\quad$ 0.51 $\quad$ & $\quad$ 0.13 $\quad$ & $\quad$ 0.27 $\quad$ & $\quad$ 0.19 $\quad$\\
& & & & \\
$ \quad\quad\quad\log$([O I]/H$\alpha$) $\quad$ &	$\quad$ 0.53 $\quad$ & $\quad$ 0.41 $\quad$ & $\quad$  0.46 $\quad$ & $\quad$ 0.61 $\quad$\\
& & & & \\
$ \quad\quad\quad\log$([O III]/H$\beta$) $\quad$ &	$\quad$ 0.45 $\quad$ & $\quad$ 0.49 $\quad$ & $\quad$ 0.51 $\quad$ & $\quad$ 0.35 $\quad$\\
\hline
\end{tabularx}
\end{table}

\section[\appendixname~\thesection]{Appendix}
In this section the complete list of the sources included in the study. The list of the sources (Table~\ref{taba2}) contains informations for the BASS sample from Koss et al. \cite{Koss}. Table~\ref{taba3}  contain the list of the fluxes of the fitted lines, used to produce the plots. 

\newpage
\newgeometry{top=2cm,bottom=2cm,left=2.5cm,right=2.5cm,heightrounded}
\begin{landscape}

\begin{table}[h!] 
\caption{The horizontal lines divide the BASS and the 4FGL. In the last row the * symbol refers to the sum of six different spectra to produce the studied one: 2016-02-06, 2016-08-12, 2016-12-08, 2016-12-29, 2017-01-20, and 2017-12-18. Continue in the next page. \label{taba2}}
\newcolumntype{C}{>{\centering\arraybackslash}X}
\begin{tabularx}{\linewidth}{p{2.5cm}p{5.5cm}p{2cm}p{2cm}p{1.5cm}p{2.5cm}p{2cm}p{2cm}p{3cm}}
\hline
\textbf{IAU name} & \textbf{Alias} & \textbf{RA} & \textbf{DEC} & \textbf{z} & \textbf{N$_H$} & \textbf{Telescope} & \textbf{Exposure} & \textbf{Date}\\
\textbf{ } & \textbf{ } & \textbf{[$^{\circ}$ J2000]} & \textbf{[$^{\circ}$ J2000]} & \textbf{[$10^{-3}$]} & \textbf{[$10^{20}$ cm$^{-2}$]} & \textbf{ } & \textbf{[ks]} & \textbf{}\\
\hline
\multicolumn{9}{c}{\textbf{BASS}} \\
\hline
J0149-5015 	& 2MASX J01492228-5015073 & 27.34& -50.25 & 2.99 & 1.75 & Palermo & 1.20 & 2010-11-12 \\
J0157+4715	& 2MASX J01571097+4715588 & 29.30 & 47.27 & 4.79 & 11.0 & Perkins & 2.40 & 2011-05-01 \\
J0206-0017 	& Mrk 1018 & 31.57 & -0.29 & 4.27 & 2.48 & SDSS & 2.70 & 2000-09-25 \\
J0214-0046 	& Mrk 590 & 33.64 & -767.00 & 2.63 & 2.77 & SDSS & 4.20 & 2003-01-10 \\
J0234-0847 	& NGC 985 & 38.66 & -8.79 & 4.30 & 3.48 & Palomar & 0.15 & 2013-08-13 \\
J0238-4038	& 2MASX J02384897-4038377 & 39.70 & -40.64 & 6.13 & 2.06 & Palomar & 0.15 & 2013-08-13 \\
J0312+5029 	& 2MASX J03120291+5029147 & 48.01 & 50.49 & 6.15 & 3.45 & Perkins & 2.40 & 2011-06-01 \\
J0330+0538 	& 2MASX J03305218+0538253 & 52.72 & 5.64 & 4.58 & 1.16 & UH & 1.22 & 2012-11-01 \\
J0333+3718 	& 2MASX J03331873+3718107 & 53.33 & 37.30 & 5.47 & 1.48 & Masetti & 1.80 & 2006-10-02 \\
J0423+0408 	& 2MASX J04234080+0408017 & 65.92 & 4.13 & 4.62 & 12.6& Palomar & 0.15 & 2012-10-18 \\
J0503+2300 	& LEDA 097068 & 75.74 & 23.00 & 5.81 & 23.4 & Palermo & 1.80 & 2009-01-29 \\
J0605-2754 	& 2MASX J06054896-2754398 & 91.45 & -27.91 & 8.98 & 2.64 & Masetti & 1.00 & 2009-11-30 \\
J0733+4555 	& 1RXS J073308.7+455511 & 113.29 & 45.92 & 14.15 & 7.19 & SDSS & 3.30 & 2004-11-05 \\
J0736+5846 	& Mrk 9 & 114.24 & 58.77 & 3.99 & 4.83 & Palomar & 0.15 & 2014-02-22 \\
J0742+4948 	& Mrk 79 & 115.64 & 49.81 & 2.21 & 5.43 & FAST& 0.72 & 2005-12-29 \\
J0752+1935 	& 2MASX J07521780+1935423 & 118.07 & 19.60 & 11.70 & 4.06 & SDSS & 3.30 & 2003-10-27 \\
J0803+0841 	& 2MASX J08032736+0841523 & 120.86 & 8.70 & 4.68 & 2.87 & SDSS & 4.50 & 2007-02-08 \\
J0804+0506 	& Mrk 1210 & 121.02 & 5.11 & 1.35 & 3.86 & Palomar & 0.15 & 2012-10-18 \\
J0829+4154 	& 2MASX J08294266+4154366 & 127.43 & 41.91 & 12.61 & 3.57 & SDSS & 3.50 & 2008-02-28 \\
J0832+3707 	& FBQS J083225.3+370736 & 128.11 & 37.13 & 9.20 & 3.27 & SDSS & 5.10 & 2002-01-23 \\
J0840+2949	& 4C +29.30 &	130.01 & 29.82	& 6.47 & 4.56 & SDSS & 22.89 & 2003-10-25\\
J0842+0759 	& 2MASX J08420557+0759253 & 130.52 & 7.99 & 13.37 & 5.76 & SDSS & 2.24 & 2003-11-21 \\
J0843+3549	& 2MASX J08434495+3549421 & 130.94 & 35.83 &	5.39 & 2.98 &  SDSS & 	2.10 & 2003-02-02 \\
J0904+5536 	& 2MASX J09043699+5536025 & 136.15 & 55.60 & 3.72 & 2.32 & SDSS & 9.00 & 2000-12-29 \\
J0918+1619 	& Mrk 704 & 139.61 & 16.31 & 2.95 & 2.74 & Palomar & 0.15 & 2015-02-17 \\
J0923+2255 	& MCG +04-22-042 & 140.93 & 22.91 & 3.30 & 3.60 & Palomar & 0.15 & 2015-02-17 \\
J0925+5219 	& Mrk 110 & 141.30 & 52.29 & 3.52 & 1.27 & SDSS & 4.80 & 2006-01-05 \\
\hline
\end{tabularx}
\end{table}
\end{landscape}

\newpage
\newgeometry{top=2cm,bottom=2cm,left=2.5cm,right=2.5cm,heightrounded}
\begin{landscape}
\begin{table}[h!] 
\newcolumntype{C}{>{\centering\arraybackslash}X}
\begin{tabularx}{\linewidth}{p{2.5cm}p{5.5cm}p{2cm}p{2cm}p{1.5cm}p{2.5cm}p{2cm}p{2cm}p{3cm}}
\hline
\textbf{IAU name} & \textbf{Alias} & \textbf{RA} & \textbf{DEC} & \textbf{z} & \textbf{N$_H$} & \textbf{Telescope} & \textbf{Exposure} & \textbf{Date}\\
\textbf{ } & \textbf{ } & \textbf{[$^{\circ}$ J2000]} & \textbf{[$^{\circ}$ J2000]} & \textbf{[$10^{-3}$]} & \textbf{[$10^{20}$ cm$^{-2}$]} & \textbf{ } & \textbf{[ks]} & \textbf{}\\
\hline
J0926+1245 	& Mrk 705 & 141.51 & 12.73 & 2.86 & 3.43 & SDSS & 5.10 & 2006-12-23 \\
J0935+2617 	& 2MASX J09352707+2617093 & 143.86 & 26.29 & 12.21 & 1.56 & SDSS & 13.02 & 2005-12-28 \\
J0942+2341 	& CGCG 122-055 & 145.52 & 23.69 & 2.13 & 2.42 & SDSS & 3.30 & 2005-12-31 \\
J0945+0738 	& 3C 227 & 146.94 & 7.42 & 8.60 & 2.00 & SDSS & 2.70 & 2003-03-27 \\
J0959+1302 	& NGC 3080 & 149.98 & 13.04 & 3.54 & 2.74 & SDSS & 2.70 & 2004-02-20 \\
J1001+2847 	& 3C 234.0 & 150.46 & 28.79 & 18.48 & 1.62 & SDSS & 2.52 & 2009-04-17 \\
J1023+1951 	& NGC 3227 & 155.88 & 19.87 & 0.33 & 1.86 & FAST & 0.33 & 2006-01-05 \\
J1043+1105 	& SDSS J104326.47+110524.2 & 160.86 & 11.09 & 4.77 & 2.38 & SDSS & 3.90 & 2004-04-20 \\
J1315+4424 	& UGC 08327 & 198.82 & 44.41 & 3.55 & 1.68 & SDSS & 1.92 & 2004-03-25 \\
J1445+2702 	& CGCG 164-019 & 221.40& 27.03 & 2.96 & 2.37 & Palomar & 0.15 & 2005-02-17 \\
J1508-0011 	& Mrk 1393 & 227.22 & -197.00 & 5.44 & 4.64 & SDSS & 2.70 & 2001-03-22 \\
\hline
\multicolumn{9}{c}{\textbf{4FGL}} \\
\hline
J0038-0207	& 3C 17 & 9.59 & -2.13 & 22.04 & 2.48 & SDSS & 4.50 & 2014-12-21 \\
J0324+3410 	& 1H 0323+342 & 51.17 & 34.18 & 6.29 & 11.7 & Asiago & 1.20 ($\times$6) & * \\ 
J0937+5008	& GB6 J0937+5008 & 144.30 & 50.15 & 27.55 & 1.31 & SDSS & 3.60 & 2014-12-01 \\
J0958+3224 	& 3C 232 & 149.59 & 32.40 & 53.06 & 1.57 & SDSS & 2.70 & 2018-01-24 \\
J1443+5201	& 3C 303 & 220.76 & 52.03 & 14.12 & 1.66 & SDSS & 2.10 & 2003-05-27 \\
J1516+0015	& PKS 1514+00 & 229.17 & 0.25 & 5.26 & 4.17 & SDSS & 4.50 & 2000-05-25 \\
J2007-4434 	& PKS 2004-447 & 301.98 & -44.58 & 24.00 & 2.97 & VLT & 0.15 & 2015-10-17 \\ 
J2118+0013 	& PMN J2118+0013 & 319.57 & 0.22 & 46.28 & 5.87 & SDSS & 3.60 & 2017-10-13 \\ 	
J2118-0732 	& TXS 2116-077 & 319.72 & -7.54 & 26.01 & 7.99 & SDSS & 2.70 & 2001-08-25 \\
\hline
\end{tabularx}
\end{table}
\end{landscape}

\newpage
\newgeometry{top=2cm,bottom=2cm,left=2.5cm,right=2.5cm,heightrounded}
\begin{landscape}
\begin{table}[h!] 
\caption{All the terms are expressed in the logarithmic form (e.g. $\log_{10}$(H$\beta_{tot}$)), and the units are erg s$^{-1}$ cm$^{-2}$. Continue in the next page. \label{taba3}}
\newcolumntype{C}{>{\centering\arraybackslash}X}
\begin{tabularx}{\linewidth}{p{2.2cm}p{1.24cm}p{1.24cm}p{2cm}p{2cm}p{1.7cm}p{2cm}p{1.24cm}p{1.24cm}p{2cm}p{2cm}p{2cm}}
\hline
\textbf{Name} & \textbf{H$\beta_{tot}$} & \textbf{H$\beta_b$} & \textbf{[OIII]$\lambda$4959} & \textbf{[OIII]$\lambda$5007} & \textbf{[OI]$\lambda$6300} & \textbf{[NII]$\lambda$6548} & \textbf{H$\alpha_{tot}$} & \textbf{H$\alpha_b$} & \textbf{[NII]$\lambda$6583} & \textbf{[SII]$\lambda$6717} & \textbf{[SII]$\lambda$6731}\\
\hline
\multicolumn{12}{c}{\textbf{BASS}} \\
\hline
J0149--5015   &  -11.57  &  -11.59  &  -12.20  &  -11.72  &  -12.65  &  -12.71  &  -10.77  &  -10.78  &  -12.24  &  -12.41  &  -12.40 \\
J0157+4715   &  -10.84  &  -10.87  &  -12.14  &  -11.66  &  -15.92  &  -11.77  &  -10.35  &  -10.40 &  -12.06  &  -12.28  &  -12.43 \\
J0206--0017   &  -11.33  &  -11.36  &  -11.89  &  -11.41  &  -12.61  &  -12.27  &  -10.51  &  -10.53  &  -11.80  &  -12.46  &  -12.50 \\
J0214--0046   &  -11.09  &  -11.13  &  -11.33  &  -10.84  &  -12.07  &  -12.01  &  -10.49  &  -10.52  &  -11.54  &  -12.23  &  -12.23 \\
J0234--0847   &  -10.74  &  -10.76  &  -11.32  &  -10.84  &  -11.97  &  -12.00  &  -9.79  &  -9.81  &  -11.53  &  -11.71  &  -11.71 \\
J0238--4038   &  -10.88  &  -10.91  &  -11.59  &  -11.10  &  -12.48  &  -12.64  &  -10.33  &  -10.36  &  -12.17  &  -12.83  &  -12.82 \\
J0312+5029   &  -10.59  &  -10.66  &  -10.95  &  -10.46  &  -11.68  &  -11.8  &  -9.74  &  -9.77  &  -11.33  &  -11.51  &  -11.58 \\
J0330+0538   &  -10.61  &  -10.79  &  -10.59  &  -10.11  &  -11.77  &  -11.79  &  -9.94  &  -10.13  &  -11.32  &  -11.78  &  -11.79 \\
J0333+3718   &  -10.73  &  -10.75  &  -11.21  &  -10.73  &  -12.24  &  -13.02  &  -10.24  &  -10.26  &  -11.77  &  -12.12  &  -12.14 \\
J0423+0408   &  -11.70  &  -11.99  &  -11.08  &  -10.60  &  -12.15  &  -12.27  &  -11.20  &  -11.72  &  -11.80  &  -11.93  &  -11.97 \\
J0503+2300   &  -10.26  &  -10.31  &  -10.88  &  -10.40  &  -11.97  &  -11.97  &  -9.58  &  -9.61  &  -11.49  &  -11.63  &  -11.63 \\
J0605--2754   &  -11.27  &  -11.32  &  -11.59  &  -11.11  &  -12.23  &  -11.77  &  -10.48  &  -10.52  &  -12.24  &  -12.16  &  -12.24 \\
J0733+4555   &  -10.68  &  -10.69  &  -11.94  &  -11.46  &  -13.03  &  -12.31  &  -10.18  &  -10.20  &  -11.84  &  -13.04  &  -13.03 \\
J0736+5846   &  -10.63  &  -10.64  &  -11.56  &  -11.07  &  -12.66  &  -12.44  &  -10.11  &  -10.13  &  -11.96  &  -12.49  &  -12.53 \\
J0742+4948   &  -10.31  &  -10.35  &  -10.76  &  -10.27  &  -11.62  &  -11.49  &  -9.67  &  -9.69  &  -11.02  &  -11.45  &  -11.51 \\
J0752+1935   &  -10.77  &  -10.80  &  -11.21  &  -10.73  &  -11.76  &  -12.09  &  -10.19  &  -10.22  &  -11.61  &  -11.68  &  -11.78 \\
J0803+0841   &  -11.54  &  -11.61  &  -11.78  &  -11.29  &  -12.30  &  -12.50  &  -10.55  &  -10.58  &  -12.03  &  -12.29  &  -12.34 \\
J0804+0506   &  -11.05  &  -11.22  &  -10.49  &  -10.01  &  -10.98  &  -11.44  &  -10.38  &  -10.91  &  -10.96  &  -11.40  &  -11.32 \\
J0829+4154   &  -11.31  &  -11.36  &  -11.73  &  -11.25  &  -12.11  &  -12.57  &  -10.58  &  -10.61  &  -12.10  &  -12.30  &  -12.26 \\
J0832+3707   &  -10.87  &  -10.89  &  -11.58  &  -11.09  &  -12.37  &  -12.26  &  -10.33  &  -10.35  &  -11.78  &  -12.27  &  -12.27 \\
J0840+2949   &  -12.50  &  -12.83  &  -11.93  &  -11.45  &  -12.66  &  -12.26  &  -11.73  &  -12.59  &  -11.78  &  -12.26  &  -12.37 \\
J0842+0759   &  -11.42  &  -11.49  &  -11.49  &  -11.01  &  -11.95  &  -12.33  &  -10.48  &  -10.51  &  -11.86  &  -12.09  &  -12.22 \\
J0843+3549   &  -12.02  &  -12.36  &  -11.48  &  -11.00  &  -11.99  &  -12.00  &  -11.29  &  -11.84  &  -11.53  &  -11.78  &  -11.83 \\
J0904+5536   &  -11.28  &  -11.32  &  -12.01  &  -11.53  &  -12.67  &  -12.35  &  -10.59  &  -10.64  &  -11.88  &  -12.28  &  -12.34 \\
J0918+1619   &  -10.16  &  -10.19  &  -10.94  &  -10.46  &  -12.38  &  -10.86  &  -9.45  &  -9.75  &  -11.35  &  -11.97  &  -12.05 \\
J0923+2255   &  -10.52  &  -10.60  &  -10.98  &  -10.50  &  -12.04  &  -11.32  &  -9.97  &  -10.03  &  -11.79  &  -11.92  &  -11.95 \\
J0925+5219   &  -11.11  &  -11.26  &  -11.14  &  -10.66  &  -11.70  &  -13.03  &  -10.24  &  -10.30  &  -12.56  &  -11.83  &  -11.87 \\
J0926+1245   &  -10.51  &  -10.54  &  -11.07  &  -10.59  &  -12.05  &  -11.83  &  -9.97  &  -10.00  &  -11.36  &  -11.84  &  -11.85 \\
\hline
\end{tabularx}
\end{table}
\end{landscape}

\newpage
\newgeometry{top=2cm,bottom=2cm,left=2.5cm,right=2.5cm,heightrounded}
\begin{landscape}
\begin{table}[h!] 
\newcolumntype{C}{>{\centering\arraybackslash}X}
\begin{tabularx}{\linewidth}{p{2.2cm}p{1.24cm}p{1.24cm}p{2cm}p{2cm}p{1.7cm}p{2cm}p{1.24cm}p{1.24cm}p{2cm}p{2cm}p{2cm}}
\hline
\textbf{Name} & \textbf{H$\beta_{tot}$} & \textbf{H$\beta_b$} & \textbf{[OIII]$\lambda$4959} & \textbf{[OIII]$\lambda$5007} & \textbf{[OI]$\lambda$6300} & \textbf{[NII]$\lambda$6548} & \textbf{H$\alpha_n$} & \textbf{H$\alpha_b$} & \textbf{[NII]$\lambda$6583} & \textbf{[SII]$\lambda$6717} & \textbf{[SII]$\lambda$6731}\\
\hline
J0935+2617   &  -11.28  &  -11.30  &  -11.99  &  -11.51  &  -12.71  &  -12.99  &  -10.45  &  -10.46  &  -12.52  &  -12.77  &  -12.86 \\
J0942+2341   &  -11.41  &  -11.45  &  -11.85  &  -11.37  &  -12.47  &  -12.23  &  -10.77  &  -10.83  &  -11.76  &  -12.35  &  -12.34 \\
J0945+0738   &  -11.01  &  -11.02  &  -11.75  &  -11.27  &  -12.92  &  -13.18  &  -10.35  &  -10.36  &  -12.71  &  -12.68  &  -12.77 \\
J0959+1302   &  -11.07  &  -11.08  &  -12.45  &  -11.63  &  -16.52  &  -12.73  &  -10.57  &  -10.58  &  -12.26  &  -12.95  &  -12.88 \\
J1001+2847   &  -11.68  &  -12.2  &  -11.21  &  -10.73  &  -12.60  &  -12.36  &  -10.98  &  -11.28  &  -11.88  &  -12.48  &  -12.47 \\
J1023+1951   &  -10.24  &  -10.30  &  -10.50  &  -10.02  &  -11.06  &  -10.18  &  -9.58  &  -9.66  &  -10.54  &  -10.84  &  -10.85 \\
J1043+1105   &  -11.11  &  -11.18  &  -11.50  &  -11.02  &  -12.29  &  -12.77  &  -10.37  &  -10.42  &  -12.29  &  -12.41  &  -12.48 \\
J1315+4424   &  -11.75  &  -12.30  &  -11.90  &  -11.42  &  -11.79  &  -11.63  &  -11.03  &  -11.85  &  -11.35  &  -11.45  &  -11.54 \\
J1445+2702   &  -11.27  &  -11.54  &  -10.94  &  -10.45  &  -12.34  &  -11.92  &  -10.73  &  -10.99  &  -11.44  &  -12.04  &  -12.04 \\
J1508--0011   &  -11.78  &  -12.47  &  -11.18  &  -10.69  &  -12.09  &  -11.75  &  -10.98  &  -11.29  &  -11.28  &  -11.86  &  -11.86 \\

\hline
\multicolumn{12}{c}{\textbf{4FGL}} \\
\hline
J0038--0207   &  -12.56  &  -12.70  &  -12.62  &  -12.14  &  -12.64  &  -13.47  &  -11.87  &  -11.95  &  -13.00  &  -13.46  &  -13.33 \\
J0324+3410   &  -10.95  &  -10.96  &  -12.25  &  -11.76  &  -13.24  &  -12.71  &  -10.59  &  -10.61  &  -12.23  &  -12.52  &  -12.59 \\
J0937+5008   &  -12.88  &  -12.95  &  -13.39  &  -12.91  &  -13.72  &  -14.54  &  -12.30  &  -12.39  &  -14.06  &  -13.31  &  -13.41 \\
J0958+3224   &  -11.90  &  -12.94  &  -12.00  &  -11.52  &  -13.14  &  -13.05  &  -11.19  &  -11.21  &  -12.58  &  -12.77  &  -12.92 \\
J1443+5201   &  -12.07  &  -12.19  &  -12.66  &  -12.18  &  -12.55  &  -13.45  &  -11.33  &  -11.39  &  -12.98  &  -13.04  &  -13.00 \\
J1516+0015   &  -12.08  &  -12.29  &  -12.19  &  -11.71  &  -12.00  &  -12.47  &  -11.39  &  -11.50  &  -12.00  &  -12.18  &  -12.26 \\
J2007--4434   &  -12.65  &  -12.70  &  -12.96  &  -12.48  &  -13.35  &  -13.54  &  -11.89  &  -11.94  &  -13.07  &  -13.12  &  -13.16 \\
J2118+0013   &  -12.77  &  -12.81  &  -13.07  &  -12.59  &  -13.70  &  -14.00  &  -12.16  &  -12.21  &  -13.53  &  -13.67  &  -13.78 \\
J2118--0732   &  -12.60  &  -12.91  &  -13.48  &  -13.00  &  -13.20  &  -13.45  &  -11.82  &  -12.02  &  -12.98  &  -13.94  &  -13.99 \\
\hline
\end{tabularx}
\end{table}
\end{landscape}

\newpage
\restoregeometry

\end{document}